\documentclass[10pt]{article}

\usepackage{amsthm,amsfonts,amsmath,amssymb}
\usepackage{pxfonts}
\usepackage{euscript}
\usepackage{hyperref}

\usepackage{xypic}
\usepackage{graphicx}
\usepackage{epstopdf} 

\newcommand\void[1]       {}

\newtheorem{thm}{Theorem}

\newtheorem{prop-ph}[thm]{Proposition$^{\mathrm{ph}}$}
\newtheorem{cor-ph}[thm]{Corollary$^{\mathrm{ph}}$}
\newtheorem{lemma-ph}[thm]{Lemma$^{\mathrm{ph}}$}
\newtheorem{thm-ph}[thm]{Theorem}

\theoremstyle{definition}

\newtheorem{rema}[thm]{Remark}
\newtheorem{expl}[thm]{Example}

\newtheorem{defn-ph}[thm]{Definition$^{\mathrm{ph}}$}

\numberwithin{equation}{section}
\numberwithin{thm}{section}

\newcommand\be            {\begin{equation}}
\newcommand\ee            {\end{equation}}
\newcommand\bea           {\begin{eqnarray}}
\newcommand\eea         {\end{eqnarray}}
\newcommand\bnu          {\begin{enumerate}}
\newcommand\enu          {\end{enumerate}}

\newcommand\TO           {\EuScript{TO}}

\newcommand\id            {\mathrm{id}}

\newcommand\fun     {\EuScript{F}\mathrm{un}}

\newcommand\one    {\mathbf{1}}

\newcommand{\bulk}   {\underline{\sl bulk} }

\newcommand{\pf}{\begin{proof}}
\newcommand{\epf}{\end{proof}}

\newcommand\Cb            {\mathbb{C}}

\newcommand\Rb            {\mathbb{R}}

\newcommand\EA           {\EuScript{A}}
\newcommand\EB           {\EuScript{B}}
\newcommand\EC           {\EuScript{C}}
\newcommand\ED           {\EuScript{D}}
\newcommand\EE          {\EuScript{E}}
\newcommand\EF          {\EuScript{F}}

\newcommand\EM          {\EuScript{M}}
\newcommand\EN         {\EuScript{N}}
\newcommand\EO         {\EuScript{O}}

\newcommand\EX         {\EuScript{X}}
\newcommand\EY         {\EuScript{Y}}

\newcommand\Z           {\mathfrak{Z}}

\newcommand\FZ           {\mathfrak{Z}}

\newcommand\bk    {\mathbf{H}}

\begin{document}

\begin{center} \LARGE
Boundary-bulk relation in topological orders
\\
\end{center}

\vskip 2em
\begin{center}
{\large
Liang Kong$^{a}$,\,
Xiao-Gang Wen$^{b}$,
Hao Zheng$^{c}$\,
~\footnote{Emails:
{\tt  kong.fan.liang@gmail.com, xgwen@mit.edu, hzheng@math.pku.edu.cn}}}
\\[2em]
$^a$ Department of Mathematics and Statistics\\
University of New Hampshire, Durham, NH 03824, USA
\\[1em]
$^b$ Department of Physics, Massachusetts Institute of Technology, \\
Cambridge, Massachusetts 02139, USA
\\[1em]
$^c$ Department of Mathematics, Peking University,\\
Beijing, 100871, China
\end{center}

\vskip 3em

\begin{abstract}
In this paper, we study the relation between an anomaly-free $n+$1D topological
order, which are often called $n+$1D topological order in physics literature,
and its $n$D gapped boundary phases. We argue that the $n+$1D bulk anomaly-free
topological order for a given $n$D gapped boundary phase is unique.  This
uniqueness defines the notion of the ``\bulk'' for a given gapped boundary
phase.  In this paper, we show that the $n+$1D ``\bulk'' phase is given by the
``center'' of the $n$D boundary phase.  In other words, the geometric notion of
the ``\bulk'' corresponds precisely to the algebraic notion of the ``center''.
We achieve this by first introducing the notion of a morphism between two
(potentially anomalous) topological orders of the same dimension, then proving
that the notion of the ``\bulk'' satisfies the same universal property as that
of the ``center'' of an algebra in mathematics, i.e.  ``\bulk = center''.  The
entire argument does not require us to know the precise mathematical
description of a (potentially anomalous) topological order. This result leads
to concrete physical predictions.
\end{abstract}



\section{Introduction}


Topological orders have attracted a lot of attention in recent years among condensed matter physicists because it is a new kind of order beyond Landau's symmetry breaking theory (see reviews \cite{wen,nssfs}). There are a lot of attempts at defining the notion of a topological order at the physics level of rigor \cite{CGW,ZW}. A mathematically rigorous definition is still out of reach.

\medskip
An $n+$1D (spacetime dimension) topological order is called {\it anomaly-free} if it can be realized by an $n+$1D lattice model; and is called {\it anomalous} if otherwise \cite{kong-wen}. In this work, we are only interested in $n+$1D anomaly-free topological orders that allow gapped $n$D boundaries, which are in general not unique. If the $n+$1D bulk phase is not trivial, then the boundary phases are {\it anomalous $n$D topological orders}. In this work, when we refer to both anomaly-free and anomalous topological orders, we use the term a {\it potentially anomalous topological order}, or simply a {\it topological order}.
The main goal of this paper is to give a precise description of the relation
between an anomaly-free $n+$1D topological order and its gapped $n$D boundary
phases.

\medskip
When an anomaly-free 2+1D topological order admits gapped boundaries, it is completely determined by its topological excitations (see reviews \cite{kitaev2}). The 1+1D gapped boundary phases are also determined by its topological excitations. The topological excitations on gapped 1+1D boundaries were first studied in the 2+1D toric code model \cite{kitaev1} by Bravyi and Kitaev in \cite{bravyi-kitaev}. It was later generalized to Levin-Wen models \cite{lw-mod} with gapped boundaries in \cite{kitaev-kong}, where the topological excitations on the boundary of such a lattice model were shown to
form a unitary fusion category $\EC$, and those in the bulk form a unitary
modular tensor category which is given by the Drinfeld center $Z(\EC)$ of
$\EC$ (see also \cite{lan-wen}).  But these works did not address the
uniqueness of the bulk phase for a given boundary. In \cite{fsv}, it was shown
model-independently that among all possible bulk phases associated to the same
boundary phase $\EC$, the Drinfeld center $Z(\EC)$ is the universal one (a
terminal object).  One way to complete the proof of the uniqueness of the bulk
is to view the gapped boundary as a consequence of anyon condensation \cite{bs}
of a given bulk theory $\ED$ to the trivial phase. This idea leads to a
classification of gapped boundaries for abelian 2+1D topological theories
\cite{kapustin-saulina,WW1263,levin,bjq}. This result, together with results in
\cite{fsv}, implies the uniqueness of the bulk for abelian 2+1D topological
theories. The proof for general 2+1D topological orders appeared in the
mathematical theory of anyon condensation developed in \cite{kong-anyon}, in
which it was shown that such a condensation is determined by a Lagrangian
algebra $A$ in $\ED$, and $\EC$ is monoidally equivalent to the category
$\ED_A$ of $A$-modules in $\ED$. Moreover, we have $Z(\ED_A)\simeq \ED$. This
completes the proof of the {\it bulk-boundary relation} in 2+1D, which says that
the 2+1D bulk phase $\ED$ for a given 1+1D boundary $\EC$ is unique, and is
given by the Drinfeld center of $\EC$, i.e. bulk = center for simplicity.

\smallskip
Does this bulk-boundary relation (i.e. bulk = center) hold in higher dimensions?
In this work, we propose that the answer is yes, and provide a formal proof of this relation under some natural assumptions. There are three key steps in this formal proof:
\bnu

\item For any given $n$D potentially anomalous topological order $\EC_n$, we assume that there is a unique anomaly-free $n+$1D topological order, denoted by $\Z_n(\EC_n)$, such that $\EC_n$ can be realized as a gapped boundary of $\Z_n(\EC_n)$ (called {\it unique-bulk hypothesis}, see Sec.\,\ref{sec:ubh}). We will refer to $\Z_n(\EC_n)$ as the \bulk of $\EC_n$. Moreover, by restricting the $n+$1D topological order $\Z_n(\EC_n)$ to a 1-codimensional subspace, we obtain an $n$D topological order, denoted by $P_n(\Z_n(\EC_n))$. We assume that $P_n(\Z_n(\EC_n))$ contains all topological excitations in $\Z_n(\EC_n)$ (a consequence of the {\it self-detection hypothesis}, see Sec.\,\ref{sec:ubh}).


\item Although we do not know how to define a topological order rigorously, assuming the existence of such a definition and using the notion of the \bulk, we can define the notion of a morphism between two topological orders of the same dimension (see Sec.\,\ref{sec:morphism}). In particular, we show that there is a canonical morphism $\rho: P_n(\Z_n(\EC_n)) \boxtimes \EC_n \to \EC_n$, where $\boxtimes$ denotes the stacking of two topological orders of the same dimension.

\item We show that the pair $(P_n(\Z_n(\EC_n)), \rho)$, satisfies the universal property of the center of an algebra in mathematics (see Theorem \ref{thm:universal}). This implies that \bulk = center.

\enu
This result is independent of how we describe the boundary/\bulk phase mathematically.
It is a non-trivial result that leads to concrete physical predictions (see Remark \ref{rema:prediction}).

\medskip
We denote $n+$1D topological orders by $\EA_{n+1}, \EB_{n+1}, \EC_{n+1}, \ED_{n+1}$, etc. If the spacetime dimension $n$+1 is clear from the context, we abbreviate $\EC_{n+1}$ as $\EC$. We denote the trivial $n+$1D topological order by $\one_{n+1}$. In physics, the trivial topological order $\one_{n+1}$ corresponds to the equivalence class of the product states \cite{CGW}.

\medskip
This paper is written for working condensed matter physicists, especially for those working in the field of topological phases of matters. In order to convey the simply idea, we try to keep the categorical language to the minimum. In particular, we collect all mathematically technical parts in Remarks and Examples. The main text should be readable to those who do not have any extra background in category theory beyond those basics that have already been widely used in condensed matter physics (see \cite{kitaev2}).


\bigskip
\noindent {\bf Acknowledgement}:
X-G.W is supported by NSF Grant No.  DMR-1506475 and NSFC 11274192. HZ is supported by NSFC under Grant No. 11131008.

\section{Basics of topological orders} \label{sec:ubh}

In this section, we recall some basic facts about topological orders, state our key assumptions and set our notations.


A potentially anomalous $n$D topological order $\EC_n$ can always be realized
as a gapped boundary of an anomaly-free $n+$1D topological order $\EE_{n+1}$
\cite{kong-wen}. In physics, it is generally believed that the bulk anomaly-free topological order $\EE_{n+1}$ is uniquely determined by its gapped boundary phase $\EC_n$. One way to see this is to note that there is no preferred length scale. In order to define the boundary phase $\EC_n$, we need define the equivalence class of quantum states by allowing proper deformation of the state in an arbitrary large neighborhood of the boundary (without closing the gap). As a consequence, $\EE_{n+1}$ should be unique (see \cite{kong-wen} for details). This uniqueness is the key assumption of this paper. We highlight it here.
\begin{quote}
{\bf Unique-bulk hypothesis}: for any given $n$D potentially anomalous topological order $\EC_n$, there is a unique anomaly-free $n+$1D topological order, denoted by $\Z_n(\EC_n)$, such that $\EC_n$ can be realized as a gapped boundary of $\Z_n(\EC_n)$.
\end{quote}
We denote $\EE_{n+1}$ by $\Z_n(\EC_n)$ and refer to it as the \bulk of $\EC_n$.
It is clear that $\Z_n(\one_n)=\one_{n+1}$. In this work, we often use the following picture:
$$
\raisebox{-10pt}{
\begin{picture}(140, 25)
   \put(25,5){\scalebox{2}{\includegraphics{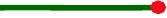}}}
   \put(0,5){
     \setlength{\unitlength}{.75pt}\put(-18,-70){
     \put(60,87)      {\footnotesize $\Z_n(\EC_n)$}
     \put(170, 86)     {\footnotesize $\EC_n$}
     }\setlength{\unitlength}{1pt}}
  \end{picture}}
$$
to illustrate the geometric relation between the boundary phase $\EC_n$ and the \bulk phase $\Z_n(\EC_n)$. More precisely, the $n+$1D \bulk phase $\Z_n(\EC_n)$ (defined on an open $n$-disk as the space manifold) is depicted by an open interval and the $n$D boundary phase $\EC_n$ (on an open $n-$1-disk) is depicted by one of the end point of the open interval.

\begin{rema} \label{rem:open-disk}
We want to remark that the notion of $n+$1D phase of matter is a local concept which is defined only on an open $n$-disk (as the space manifold). Similarly, the boundary phase is also a local concept.
Strictly speaking, the boundary phase should be defined on a neighborhood of the boundary in the bulk. Namely, the space manifold for the boundary phase is $[0,\epsilon)\times D^{n-1}$, where $D^{n-1}$ is an open $n-$1-disk. For example, when $n=2$, the bulk 2+1D phase, defined on an open 2-disk, can be described by a unitary modular tensor category together with a central charge $c\in \Rb$. Its 1+1D boundary (together with an open neighborhood in the bulk) can support a few different boundary phases, each of which lives on an open 1-disk as the space manifold. Different boundary phases are separated by higher codimensional defects \cite{kitaev-kong,ccbcn}. A boundary phase can be transformed to another boundary phase via a pure boundary phase transition without altering the bulk phase \cite{pmn}.
\end{rema}

\void{
This conclusion is based on the following believes:
\bnu
\item
 Two anomaly-free topological orders are the same iff they can join
smoothly.
\item
 Two anomaly-free topological orders can join smoothly iff their boundary
topological order, $\EC_\text{bndry}$, is the same.  Since joining two
boundaries is like stacking $\EC_\text{bdry}$ with $\bar\EC_\text{bdry}$ (the
time reversal conjugate of $\EC_\text{bdry}$.) It is believed that
$\EC_\text{bdry}$ and $\bar\EC_\text{bdry}$ can annihilate each other via anyon
condensation.
\enu
So the notion of \bulk is well defined at physics level.

Another way to see that
boundary uniquely determines that bulk is to note that
there is no preferred length scale. In order to define the boundary phase $\EC_n$ as an equivalence class of lattice models, we need allow deformation of the lattice model without closing the gap in an arbitrary large neighborhood of the boundary. As a consequence, $\EE_{n+1}$ should be unique (see \cite{kong-wen} for more discussion). We denote $\EE_{n+1}$ by $\Z_n(\EC_n)$ and refer to it as the \bulk of $\EC_n$. It is clear that $\Z_n(\one_n)=\one_{n+1}$.
}

It is well-known that topological excitations in an anomaly-free 2+1D topological order $\EC_3$ are all particle-like (of codimension 2), and they form a unitary modular tensor category, still denoted by $\EC_3$ (see for example \cite{kitaev2}). The only 1-codimensional defect (or domain wall) in $\EC_3$ is the trivial one. By restricting to the trivial 1-codimensional domain wall, all the particles have to fuse along the wall. No braiding structure remains on the wall. Therefore, restricting $\EC_3$ to the trivial 1-codimensional domain wall, we obtain a 1+1D topological order, denoted by $P_2(\EC_3)$, which is given by the same unitary fusion category as $\EC_3$ but forgetting its braiding structure. We would like to generalize this fact to all dimensions under the following assumption:
\begin{quote}
{\bf Self-detection hypothesis}: all topological excitations in an anomaly-free topological order $\EC_{n+1}$ should be able to detect themselves via double braidings \cite{kong-wen, kong-wen-zheng}.
\end{quote}
As a consequence, $\EC_{n+1}$ can not contain any non-trivial topological excitations (or defects) of codimension 1 because two is the smallest codimension for an excitation to be braided with another excitation. Therefore, by restricting the topological order $\EC_{n+1}$ to the trivial excitation of codimension 1 (i.e. the trivial domain wall), we obtain an $n$D topological order $P_n(\EC_{n+1})$. In this restricting process, we do not lose any non-trivial topological excitations, nor any information of the fusion among them in $n$ spatial dimensions, but only forget the information of their fusion in the $n$+1th direction, which further encodes the braidings among excitations of codimension 2. Moreover, by double folding the anomaly-free topological order $\EC_{n+1}$, we create a double layered system $\EC_{n+1}\boxtimes \overline{\EC}_{n+1}$ with a gapped boundary phase
$P_n(\EC_{n+1})$ (i.e. Eq.\,(\ref{eq:ZPC})), where $\boxtimes$ is the stacking operation explained below and $\overline{\EC}_{n+1}$ is the time reverse of $\EC_{n+1}$ (because the orientation or the normal (or the time) direction of one of the two layers is flipped).

\medskip
One can stack an $n$D potentially anomalous topological order $\EA_n$ on the top of the another one $\EB_n$ without introducing any coupling between the two layers as shown in (\ref{eq:AxB1}). This operation is denoted by $\boxtimes$. More precisely, the dashed box, when viewed from far away, can be viewed as a single $n$D topological order $\EA_n\boxtimes \EB_n$ with a single (but two-layer) bulk phase $\FZ_n(\EA_n)\boxtimes \FZ_n(\EB_n)$.
\be \label{eq:AxB1}
\raisebox{-3.2em}{\begin{picture}(90, 87)
   \put(-10,0){\scalebox{2.5}{\includegraphics{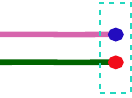}}}
   \put(-10,0){
     \setlength{\unitlength}{.75pt}\put(-18,-40){
     \put(123, 55)     {\small $\EB_n$}
     \put(123, 112)     {\small $\EA_n$}
     \put(110,140)   {\small $\EA_n \boxtimes \EB_n$}
     \put(30,108)    {\small $\FZ_n(\EA_n)$}
     \put(30,58)     {\small $\FZ_n(\EB_n)$}
     }\setlength{\unitlength}{1pt}}
  \end{picture}}
\quad\quad\quad=
\raisebox{-10pt}{
\begin{picture}(140, 25)
   \put(25,5){\scalebox{2}{\includegraphics{pic-CZC-eps-converted-to.pdf}}}
   \put(0,5){
     \setlength{\unitlength}{.75pt}\put(-18,-70){
     \put(60,88)      {\footnotesize $\FZ_n(\EA_n)\boxtimes \FZ_n(\EB_n)$}
     \put(160, 88)     {\footnotesize $\EA_n \boxtimes \EB_n$}
     }\setlength{\unitlength}{1pt}}
  \end{picture}}
\ee
This stacking operation is completely symmetric. It does not matter if we put $\EA_n$ on the top or the bottom of $\EB_n$ because there is no coupling between them. The resulting new phase is the same, i.e.
\be \label{eq:symmetric}
\EA_n\boxtimes\EB_n = \EB_n\boxtimes\EA_n.
\ee
Clearly, we have $\one_n \boxtimes \EA_n = \EA_n$. Since the \bulk is unique, we should also have the following identity:
$$
\Z_n(\EA_n\boxtimes \EB_n) = \Z_n(\EA_n) \boxtimes \Z_n(\EB_n).
$$

More generally, we can glue $\EA_n$ with $\EB_n$ by a potentially anomalous $n+$1D phase $\EC_{n+1}$ to obtain a new $n$D topological order, denoted by $\EA_n\boxtimes_{\EC_{n+1}} \EB_n$, as shown in (\ref{eq:AxB2}). More precisely, the dashed box, when viewed from far away, can be viewed as a single $n$D topological order $\EA_n \boxtimes_{\EC_{n+1}} \EB_n$, which has a single bulk phase given by $\EA_n' \boxtimes_{\FZ_{n+1}(\EC_{n+1})} \EB_n'$.\footnote{In (\ref{eq:AxB2}), note that $\EA_{n+1}'\neq \Z_n(\EA_n)$ and $\EB_{n+1}'\neq \Z_n(\EB_n)$. Actually, by the uniqueness of the \bulk, as a generalization of Eq.\,(\ref{eq:ZPC}), we have $\Z_n(\EA_n) = \overline{\EC}_{n+1} \boxtimes_{\Z_{n+1}(\EC_{n+1})}\EA_{n+1}'$.}
It is clear that $\boxtimes=\boxtimes_{\one_{n+1}}$. This operation $\boxtimes_{\EC_{n+1}}$ is not symmetric in general\footnote{But we can rotate the left picture in (\ref{eq:AxB2}) around a horizontal line pass through the middle point of $\EC_{n+1}$ by 180 degrees. We obtain the same boundary phase $\EA_n \boxtimes_{\EC_{n+1}} \EB_n=\EB_n\boxtimes_{\overline{\EC_{n+1}}} \EA_n$, where $\overline{\EC_{n+1}}$ is the mirror reflection of $\EC_{n+1}$ along the same line and is not equivalent to $\EC_{n+1}$ in general. This also explains Eq.\,(\ref{eq:symmetric}) because $\one_{n+1}=\overline{\one_{n+1}}$.}.
\be \label{eq:AxB2}
\raisebox{-3.4em}{\begin{picture}(115, 87)
   \put(0,0){\scalebox{2.5}{\includegraphics{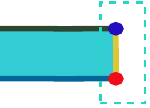}}}
   \put(0,0){
     \setlength{\unitlength}{.75pt}\put(-18,-40){
     \put(134, 89)     {\small $ \EC_{n+1} $}
     \put(125, 50)     {\small $\EB_n$}
     \put(125, 125)     {\small $\EA_n$}
     \put(110,149)   {\small  $\EA_n \boxtimes_{\EC_{n+1}} \EB_n$}
     \put(30,126)    {\small $\EA_{n+1}'$}
     \put(30,53)     {\small $\EB_{n+1}'$}
     \put(30,89)    {\small $\Z_{n+1}(\EC_{n+1})$}
     }\setlength{\unitlength}{1pt}}
  \end{picture}}
\quad\quad=
\raisebox{-10pt}{
\begin{picture}(140, 25)
   \put(25,5){\scalebox{2}{\includegraphics{pic-CZC-eps-converted-to.pdf}}}
   \put(0,5){
     \setlength{\unitlength}{.75pt}\put(-18,-70){
     \put(60,88)      {\footnotesize $\EA_n' \boxtimes_{\FZ_{n+1}(\EC_{n+1})} \EB_n'$}
     \put(160, 88)     {\footnotesize $\EA_n \boxtimes_{\EC_{n+1}} \EB_n$}
     }\setlength{\unitlength}{1pt}}
  \end{picture}}
\ee


\medskip
A gapped domain wall (or a wall) $\EM_n$ between two anomaly-free\footnote{A more general notion of a (potentially anomalous) gapped domain wall between two potentially anomalous topological orders can be introduced (see \cite[Sec.\,6.1]{kong-wen-zheng}). But we do not need it in this work.} $n+$1D topological orders $\EC_{n+1}$ and $\ED_{n+1}$ is itself a potentially anomalous $n$D topological order. Moreover, we have
\be
\Z_n(\EM_n) = \EC_{n+1}\boxtimes \overline{\ED}_{n+1},
\ee
where $\overline{\ED}_{n+1}$ is the time reverse of $\ED_{n+1}$. As a special case, we have
\be \label{eq:ZPC}
\Z_n(P_n(\EC_{n+1}))=\EC_{n+1}\boxtimes \overline{\EC}_{n+1}.
\ee
An $n$D topological order $\EA_n$ can be viewed as a wall between $\Z_n(\EA_n)$ and $\one_{n+1}$.

A wall $\EM_n$ between $\EC_{n+1}$ and $\ED_{n+1}$ can be fused with a wall $\EN_n$ between $\ED_{n+1}$ and $\EE_{n+1}$ to obtain a wall $\EM_n\boxtimes_{\ED_{n+1}} \EN_n$ between $\EC_{n+1}$ and $\EE_{n+1}$. This fusion operation of walls is clearly associative, i.e. for a wall $\EO_n$ between $\EE_{n+1}$ and $\EF_{n+1}$,
\be
(\EM_n \boxtimes_{\ED_{n+1}} \EN_n) \boxtimes_{\EE_{n+1}} \EO_n
= \EM_n \boxtimes_{\ED_{n+1}} (\EN_n \boxtimes_{\EE_{n+1}} \EO_n). \label{eq:boxtimes-associative}
\ee
For simplicity, we denote the two sides of Eq.\,(\ref{eq:boxtimes-associative}) by $\EM \boxtimes_\ED \EN \boxtimes_\EE \EO$ in the rest of this paper. We have the following identities:
\be  \label{eq:BCD}
P_n(\ED_{n+1}) \boxtimes_{\ED_{n+1}} \EN_n = \EN_n = \EN_n \boxtimes_{\EE_{n+1}} P_n(\EE_{n+1}).
\ee

\begin{figure}
$$
 \raisebox{-50pt}{
  \begin{picture}(100,100)
   \put(50,8){\scalebox{1.5}{\includegraphics{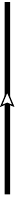}}}
   \put(50,8){
     \setlength{\unitlength}{.75pt}\put(-151,-235){
     \put(110,320)  {\scriptsize $ \EC_{n+1} $}
     \put(185,320)  {\scriptsize $ \ED_{n+1} $}
     \put(148,355)  {\scriptsize $ \EM_n $}
     \put(148,225)  {\scriptsize $\EM_n $}
     \put(110, 290) {\scriptsize $\times$}
     }\setlength{\unitlength}{1pt}}
  \end{picture}}
  \rightsquigarrow
   \raisebox{-50pt}{
   \begin{picture}(130,100)
   \put(50,8){\scalebox{1.5}{\includegraphics{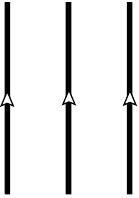}}}
   \put(50,8){
     \setlength{\unitlength}{.75pt}\put(-151,-235){
     \put(120,320)  {\scriptsize $ \EC_{n+1} $}
     \put(198, 320)  {\scriptsize $\EC_{n+1}$}
     \put(163,320)  {\scriptsize $\ED_{n+1}$}
     \put(240, 320)  {\scriptsize $\ED_{n+1}$}
     \put(148,355)  {\scriptsize $ \EM_n $}
     \put(183,355)  {\scriptsize $\EN_n $}
     \put(220,355)  {\scriptsize $ \EM_n $}
     \put(203, 290) {\scriptsize $\times$}
     }\setlength{\unitlength}{1pt}}
  \end{picture}}
  \rightsquigarrow
  \raisebox{-50pt}{
   \begin{picture}(130,100)
   \put(50,8){\scalebox{1.5}{\includegraphics{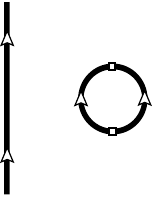}}}
   \put(50,8){
     \setlength{\unitlength}{.75pt}\put(-151,-235){
     \put(120,290)  {\scriptsize $ \EC_{n+1} $}
     \put(170,330)  {\scriptsize $ \ED_{n+1} $}
     \put(170,250)  {\scriptsize $ \ED_{n+1} $}
     \put(148,355)  {\scriptsize $ \EM_n $}
     \put(148,225)  {\scriptsize $\EM_n $}
     \put(242,288) {\scriptsize $\EM_n $}
     \put(178,288) {\scriptsize $\EN_n $}
     \put(214, 288) {\scriptsize $\times$}
     }\setlength{\unitlength}{1pt}}
  \end{picture}}
$$
\caption{$\EM_n$ is an invertible domain wall between two anomaly-free $n+$1D topological orders $\EC_{n+1}$ and $\ED_{n+1}$, and $\EN_n$ is its inverse. The ``$\times$" in these pictures represents a non-trivial topological excitation in the $\EC_{n+1}$-phase. Note that the non-trivialness requires ``$\times$" to be at least 2-codimensional. These pictures depict a process of the excitation ``$\times$" tunneling through the $\EM_n$ wall. In particular, the second ``$\rightsquigarrow$'' is obtained by annihilating $\EN_n$ with the $\EM_n$ on the right side of $\EN_n$.
Similarly, there is a tunneling process from $\ED_{n+1}$ to $\EC_{n+1}$, which is inverse to it. This gives a way to identify the $\EC_{n+1}$-phase with the $\ED_{n+1}$-phase.
}
\label{fig:tunneling}
\end{figure}

A gapped domain wall $\EM_n$ between two anomaly-free $\EC_{n+1}$ and $\ED_{n+1}$ is called {\it invertible} if there is a gapped domain wall $\EN_n$ between $\ED_{n+1}$ and $\EC_{n+1}$ such that
\be \label{eqn:inv}
\EM_n \boxtimes_{\ED_{n+1}} \EN_n = P_n(\EC_{n+1}), \quad\quad
\EN_n \boxtimes_{\EC_{n+1}} \EM_n = P_n(\ED_{n+1}).
\ee
Such an invertible domain wall $\EM_n$ provides a way to identify $\EC_{n+1}$ with $\ED_{n+1}$ as depicted in Fig.\,\ref{fig:tunneling}.

\void{
\begin{figure}[tb]
\begin{center}
\includegraphics[scale=0.7]{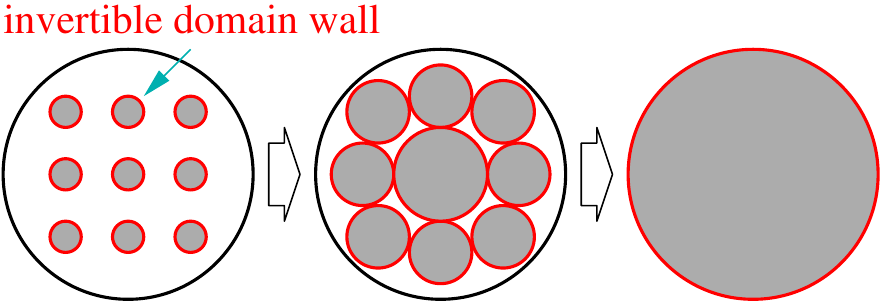}
\end{center} \label{fig:bubble}
\caption{
If the topological orders $\EC_{n+1}$ (white area) and $\ED_{n+1}$ (shaded area) are separated by an invertible domain wall (the red
line), one can deform $\EC_{n+1}$ to $\ED_{n+1}$ via local unitary transformations without closing the gap.
}
\label{bubble}
\end{figure}
Moreover, \emph{two anomaly-free topological orders $\EC_{n+1}$ and $\ED_{n+1}$ are the same, i.e. $\EC_{n+1}=\ED_{n+1}$, iff there exists an invertible domain wall $\EM_n$ between them}.
}


\begin{rema} {\rm
Gapped domain walls between topological orders in arbitrary dimensions have not been extensively studied (see some discussion in \cite{kong-wen-zheng}). They are relatively well understood in 2+1D (see \cite{kong-anyon,fsv,fs,lww,yasu1,ai,yasu2}) and in 1+1D (see \cite{kong-wen-zheng}).
}
\end{rema}

\void{
A gapped domain wall $\EM_n$ between two anomaly-free $n+$1D topological orders $\EC_{n+1}$ and $\ED_{n+1}$ is itself an anomalous $n$D topological order. Moreover, we have $\Z_n(\EM_n) = \EC_{n+1}\boxtimes \overline{\ED}_{n+1}$, where $\overline{\ED}_{n+1}$ is the time reverse of $\ED_{n+1}$. As a special case, we have
\be \label{eq:ZPC}
\Z_n(P_n(\EC_{n+1}))=\EC_{n+1}\boxtimes \overline{\EC}_{n+1}.
\ee

$$
\EM_n \boxtimes_{\ED_{n+1}} \EN_n = P_n(\EC_{n+1}), \quad\quad
\EN_n \boxtimes_{\EC_{n+1}} \EM_n = P_n(\ED_{n+1}).
$$
}

\section{The universal property of the center of an algebra} \label{sec:center}

In this section, we recall the universal property of the center of an ordinary algebra.

\medskip

An algebra $A$ over a field $k$ is a triple $(A,\; A\otimes A \xrightarrow{m} A,\; k \xrightarrow{\iota_A} A)$, where $m$ is the multiplication map and $\iota_A$ is the unit of $A$.
We also denote $\iota_A(1)=1_A$. Its center $Z(A)$ is defined to be the subalgebra:
$$
Z(A)=\{ z\in A \mid az =za, \, \forall a\in A\}.
$$
This definition is, however, very limited and not useful to us at all. A better definition of the center of an algebra is given by its {\it universal property} (see for example \cite{js}\cite[Section 6.1.4]{lurie}) which is applicable to many other types of algebras.

More precisely, let $m: Z(A)\otimes A \to A$ be the multiplication map, i.e. $m(z\otimes a)=za$ for $z\in Z(A)$ and $a\in A$. Note that $m$ defines a unital action on $A$, i.e. $m(1_{Z(A)}\otimes a)=a$. Equivalently, we have the following commutative diagram:
\be  \label{diag:ZAA1}
\xymatrix{
& Z(A) \otimes A \ar[rd]^m  & \\
k\otimes A = A \ar[rr]^{\id_A} \ar[ru]^{\iota_{Z(A)} \otimes \id_A}  & & A\,\, .
}
\ee
Moreover, $m$ is an algebra homomorphism, i.e. $m(z\otimes a)m(z'\otimes a') = m(z z'\otimes a a')$ for $z,z'\in Z(A)$ and $a,a'\in A$.

The pair $(Z(A), m)$ satisfies the following universal property:
\begin{itemize}
\item Given another pair $(X,\; X\otimes A \xrightarrow{f}A)$ where $X$ is an algebra, $f$ is a unital action and an algebra homomorphism, there is a unique algebra homomorphism $\underline{f}: X\to Z(A)$ such that $m \circ (\underline{f}\otimes \id_A) = f$,
or diagrammatically, we have the following commutative diagram:
\be  \label{diag:univ-prop-1}
\xymatrix{
& Z(A) \otimes A \ar[rd]^m  & \\
X\otimes A \ar[rr]^f \ar[ru]^{\underline{f} \otimes \id_A}  & & A\,\, .
}
\ee
\void{
\be \label{diag:univ-prop-1}
\raisebox{4em}{
\xymatrix@R=1em@C=2em{
&  & Z(A) \otimes A \ar@/^1.5pc/[rrdddd]^m &  & \\
& & &  & \\
& & X \otimes A \ar[uu]^{\underline{f} \otimes \id_A} \ar[rrdd]^f &  & \\
& &  & & \\
A \ar[uurr]^{\iota_X \otimes \id_A} \ar@/^1.5pc/[uuuurr]^{\iota_{Z(A)} \otimes \id_A}  \ar[rrrr]^{\id_A} & & & & A\, .
}}
\ee
}
\end{itemize}
Indeed, if $f: X\otimes A \to A$ is a unital action and an algebra homomorphism, then
$$f(x\otimes 1_A)a = f(x\otimes 1_A)f(1_X\otimes a) = f(x\otimes a) = f(1_X\otimes a)f(x\otimes 1_A) = af(x\otimes 1_A)$$
for all $a\in A$, where the first and the last equalities hold because $f$ is a unital action, the second and the third equalities hold because $f$ is an algebra homomorphism. Therefore, the assignment $x\mapsto f(x\otimes 1_A)$ defines an algebra homomorphism $\underline{f}: X \to Z(A)$ rendering (\ref{diag:univ-prop-1}) commutative. By restricting the diagram to the subset $X\otimes 1_A \subset X\otimes A$, we see that $\underline{f}$ is the unique map making the diagram commutative. This shows that the pair $(Z(A), m)$ satisfies the universal property. Note that in the special case $(X,f)=(Z(A),m)$, we have $\underline{f}=\id_{Z(A)}$.

\medskip
On the other hand, suppose $(Y,g)$ is another such a pair satisfying this universal property. Then $g$ induces an algebra homomorphism $\underline{g}: Y \to Z(A)$. Since $(Y,g)$ also satisfies the universal property, the map $m: Z(A)\otimes A \to A$ also induces an algebra homomorphism $\underline{m}: Z(A) \to Y$. Then $\underline{g}\circ \underline{m}$ has to be the identity map $\id_{Z(A)}$ by the uniqueness in the universal property. Similarly, $\underline{m}\circ \underline{g}=\id_Y$. Namely, $\underline{g}$ and $\underline{m}$ are inverse to each other, hence identify $Y$ with $Z(A)$. Therefore, the universal property determines $(Z(A), m)$ uniquely up to canonical isomorphism, hence provides an alternative approach to define the notion of center.

\begin{rema}
The center $Z(A)$ of an ordinary algebra $A$ is always a subalgebra. Thus the data $m$ in the pair $(Z(A),m)$ is redundant. However, this is not the case for other types of algebras naturally arising in mathematics. One should keep in mind that the center of an algebra is a pair rather merely an algebra.
\end{rema}


In mathematical language, the collection of algebras over $k$ form a symmetric monoidal category $Alg$. Roughly speaking, an algebra $A$ is referred to as an {\it object} of $Alg$. An algebra homomorphism $f:A\to B$ is referred to as a {\it morphism} between the objects $A$ and $B$. Two morphisms $f:A\to B$ and $g:B\to C$ can be composed to give a new morphism $g\circ f:A\to C$. One has an identity morphism $\id_A:A\to A$ for every object $A$. Moreover, there is a binary operation $\otimes$ on $Alg$ which carries a pair of objects $A,B$ to their tensor product $A\otimes B$. This binary operation is {\it symmetric}, i.e. $A\otimes B=B\otimes A$, and {\it unital}, i.e. there is a distinguished object $k$ such that $k\otimes A = A$ for all $A$. This symmetric monoidal category $Alg$ satisfies an additional property: there exists a unique morphism $\iota_A:k\to A$ for every object $A$. These are all the data that we have used to state the universal property of the center of an algebra. Once given such a symmetric monoidal category no matter how crazy the objects are, one is able to write down the universal property and define the notion of center.

\medskip
The collection of $n$D topological orders almost form a symmetric monoidal category. For example, there is a symmetric binary operation $\boxtimes$ (recall Eq.\,(\ref{eq:symmetric})) that carries a pair of $n$D topological orders $\EA_n,\EB_n$ to $\EA_n\boxtimes\EB_n$, and there is a distinguished $n$D topological order $\one_n$ such that $\one_n\boxtimes\EA_n=\EA_n$ for all $\EA_n$. Once we know what is a morphism between two $n$D topological orders, we can define the center of an $n$D topological orders by applying the universal property. This is the subject of the next two sections.


\section{A morphism between two topological orders} \label{sec:morphism}

Although we do not have a rigorous definition of topological order, we may treat it as a black box and use it to give a physical definition of a morphism between two topological orders.

\medskip

A {\it morphism} $f: \EC_n \to \ED_n$ between two $n$D topological orders $\EC_n$ and $\ED_n$ is a gapped domain wall $f_n$, viewed as an $n$D topological order, between two $n+$1D anomaly-free topological orders $\Z_n(\EC_n)$ and $\Z_n(\ED_n)$ such that $f_n \boxtimes_{\Z_n(\EC_n)}  \EC_n = \ED_n$.

The geometric idealization of the physical configuration associated to this morphism can be depicted as follows:
\be \label{pic:morphism}
\raisebox{-10pt}{
\begin{picture}(140, 30)
   \put(0,5){\scalebox{2}{\includegraphics{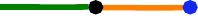}}}
   \put(-25,5){
     \setlength{\unitlength}{.75pt}\put(-18,-70){
     \put(60,87)      {\footnotesize $\Z_n(\ED_n)$}
     \put(150, 86)     {\footnotesize $\Z_n(\EC_n)$}
     \put(200, 86)     {\footnotesize $\EC_n$}
     \put(122, 88)     {\footnotesize $f_n$}
     }\setlength{\unitlength}{1pt}}
  \end{picture}}
\ee
The equality $f_n \boxtimes_{\Z_n(\EC_n)}  \EC_n = \ED_n$ means that this configuration is identical to the following one:
$$
\raisebox{-10pt}{
\begin{picture}(140, 25)
   \put(25,5){\scalebox{2}{\includegraphics{pic-CZC-eps-converted-to.pdf}}}
   \put(0,5){
     \setlength{\unitlength}{.75pt}\put(-18,-70){
     \put(60,87)      {\footnotesize $\Z_n(\ED_n)$}
     \put(170, 86)     {\footnotesize $\ED_n$}
     }\setlength{\unitlength}{1pt}}
  \end{picture}}
\quad\quad\quad
$$

\medskip

This definition is quite unconventional in either physics or mathematics. We would like to explain the intuition behind this concept. In mathematics, a morphism between two mathematical objects, such as groups, rings and algebras, are often required to preserve the internal structures of the objects. In our case, the internal structures of a topological order should include the fusing-braiding structures of the topological excitations. In order to preserve the internal structures, we realize the notion of a morphism $f: \EC_n\to \ED_n$ between two $n$D topological orders $\EC_n$ and $\ED_n$ by a physical process of ``screening'' $\EC_n$ from outside. More precisely, we glue a new $n$D topological order $f_n$ to $\EC_n$ by an $n+$1D glue $\Z_n(\EC_n)$ such that $f_n \boxtimes_{\Z_n(\EC_n)}  \EC_n = \ED_n$. Note that a topological excitation in $\EC_n$ is automatically a topological excitation in $f_n \boxtimes_{\Z_n(\EC_n)}  \EC_n = \ED_n$. In this way, the morphism $f$ supplies a map from the set of the topological excitations in $\EC_n$ to that in $\ED_n$. Intuitively, it is reasonable that this process of ``screening'' from outside should preserve the internal  structures of $\EC_n$. When $n=2$, we explain in Example\,\ref{expl:monoidal-fun} that this notion of a morphism exactly coincides with that of a unitary monoidal functor between two unitary fusion categories, which describe two 2d topological orders.


\begin{rema} {\rm
An analog of such a morphism in mathematics is an algebra homomorphism $\phi:A\to B$ between two matrix algebras $A$ and $B$. Actually, $\phi$ always factors as $A\xrightarrow{1_C\otimes\id_A}C\otimes A \simeq B$ where $C$ is another matrix algebra. Indeed, if $A$ is the algebra of $p\times p$-matrices and $B$ is the algebra of $q\times q$-matrices, then the existence of an algebra homomorphism $\phi:A\to B$ implies $p$ divides $q$ and $C$ is the algebra of $\frac q p\times\frac q p$-matrices. A miracle happens here is that an algebra homomorphism $\phi$ is encoded by an algebra $C$. This phenomenon becomes highly nontrivial in categorical settings (see Example\,\ref{expl:monoidal-fun}).
}
\end{rema}

\begin{rema} \label{rema:=}
Strictly speaking, we can not say that an $n$D topological order is equal to another unless we specify how they are identified. Such an identification can be realized by the choice of an invertible $n-$1D domain wall as we did in \cite[Def.\,4.3]{kong-wen-zheng}. In order to convey the simple idea, however, we would like to use the equality ``=" for simplicity.
\end{rema}

\void{
\begin{rema} {\rm
It is possible to make Def.\,\ref{def:morphism-2} more flexible by requiring $f_n \boxtimes_{\Z_n(\EC_n)}  \EC_n \simeq \ED_n$ as it was done in \cite[Def.\,4.3]{kong-zheng} instead of a strict equality. In order to convey the simple idea, we will only use ``='' in this work.
}
\end{rema}
}

Two morphisms $f:\EC_n \to \ED_n$ and $g:\ED_n \to \EE_n$ can be composed to get a new morphism $g \circ f : \EC_n \to \EE_n$ which is defined by the following gapped domain wall between $\Z_n(\EC_n)$ and $\Z_n(\EE_n)$:
$$
(g \circ f)_n:= g_n\boxtimes_{\Z_n(\ED_n)} f_n.
$$
The physical configuration associated to this composed morphism is depicted as follows:
$$
\raisebox{-10pt}{
\begin{picture}(140, 30)
   \put(0,5){\scalebox{2}{\includegraphics{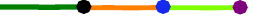}}}
   \put(-25,5){
     \setlength{\unitlength}{.75pt}\put(-18,-70){
     \put(60,87)      {\footnotesize $\Z_n(\EE_n)$}
     \put(130,87)      {\footnotesize $\Z_n(\ED_n)$}
     \put(195, 86)     {\footnotesize $\Z_n(\EC_n)$}
     \put(240, 86)     {\footnotesize $\EC_n$}
     \put(110, 88)     {\footnotesize $g_n$}
     \put(175, 88)     {\footnotesize $f_n$}
     }\setlength{\unitlength}{1pt}}
  \end{picture}}
$$

Note that this composition law is associative by Eq.\,(\ref{eq:boxtimes-associative}). That is, $h\circ(g\circ f) = (h\circ g)\circ f$ for any morphisms $f:\EC_n \to \ED_n$, $g:\ED_n \to \EE_n$ and $h:\EE_n \to \EF_n$.
Moreover, the trivial domain wall $P_n(\Z_n(\EC_n))$ between $\Z_n(\EC_n)$ and $\Z_n(\EC_n)$ defines the identity morphism $\id_{\EC_n}: \EC_n \to \EC_n$ by Eq.\,(\ref{eq:BCD}). That is, $f\circ\id_{\EC_n} = f$ for any morphism $f:\EC_n\to\ED_n$ and $\id_{\EC_n}\circ g = g$ for any morphism $g: \EE_n\to\EC_n$. A morphism $f: \EC_n \to \ED_n$ is an isomorphism (i.e. there is a morphism $g:\ED_n\to\EC_n$ such that $g\circ f=\id_{\EC_n}$ and $f\circ g=\id_{\ED_n}$) if and only if $f_n$ is an invertible domain wall (see Eq.\,(\ref{eqn:inv})).

\void{
\begin{rema}  {\rm
Recall that at the end of Sec.\,\ref{sec:ubh}, we said that an invertible $n-$1D domain wall between $\EC_n$ and $\ED_n$, when both $\EC_n$ and $\ED_n$ are anomaly-free, can be used to define the notion of an isomorphism between $\EC_n$ and $\ED_n$. It looks superficially different from Def.\,\ref{def:morphism} because we hide our choice of identification when we set two topological orders to be equal (recall Remark \ref{rema:=}). These two ways of defining isomorphisms are equivalent (see \cite[Sec.\,4.2]{kong-wen-zheng}).
}
\end{rema}
}

\begin{rema}
A morphism between two many body systems (not necessarily topological) of the same dimension can be introduced in a similar way (see \cite[Sec.\,A.3]{kong-wen-zheng}). The composition law is, however, not associative in general. For this reason, the result of this work does not apply to non-topological theories.
\end{rema}

\begin{expl} \label{exam:mor}
We give a few more examples of morphisms that will be used later. Let $\EC_n$ be an $n$D topological order.
\bnu

\item There is a unique morphism $\iota_{\EC_n}: \one_n \to \EC_n$ from $\one_n$ to $\EC_n$ which is defined by the obvious gapped domain wall $\EC_n$ between $\one_{n+1}$ and $\Z_n(\EC_n)$ as depicted below.
\be \label{eq:unit}
\raisebox{-10pt}{
\begin{picture}(140, 25)
   \put(0,5){\scalebox{2}{\includegraphics{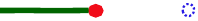}}}
   \put(-25,5){
     \setlength{\unitlength}{.75pt}\put(-18,-70){
     \put(60,87)      {\footnotesize $\Z_n(\EC_n)$}
     \put(150, 76)     {\footnotesize $\one_{n+1}$}
     \put(200, 86)     {\footnotesize $\one_n$}
     \put(122, 88)     {\footnotesize $\EC_n$}
     }\setlength{\unitlength}{1pt}}
  \end{picture}}
\ee
Note that $\EC_n \boxtimes_{\one_{n+1}} \one_n = \EC_n$.

\void{
\item There is a natural morphism $P_n(\Z_n(\EC_n)) \to \EC_n$ defined by the domain wall $P_n(\Z_n(\EC_n)) \boxtimes \EC_n$ between $\Z_n(\EC_n) \boxtimes \overline{\Z_n(\EC_n)}$ and $\Z_n(\EC_n)\boxtimes \one_{n+1}$ as depicted in the following picture:
$$
 \begin{picture}(140, 65)
   \put(20,0){\scalebox{2}{\includegraphics{pic-ZC-to-C-eps-converted-to.pdf}}}
   \put(20,0){
     \setlength{\unitlength}{.75pt}\put(-18,-19){
     \put(165, 50)       { \footnotesize $P_n(\Z_n(\EC_n))\,\, .$}
     \put(80, 37)       {\footnotesize $\EC_n$}
     \put(77, 92)     {\footnotesize $ P_n(\Z_n(\EC_n)) $}
     \put(25, 90)     {\footnotesize $ \Z_n(\EC_n) $}
     \put(125, 20)     {\footnotesize $ \overline{\Z_n(\EC_n)} $}
     \put(125, 72)   {\footnotesize $ \Z_n(\EC_n) $}
     \put(15, 34)     {\footnotesize $\one_{n+1}$ }
     }\setlength{\unitlength}{1pt}}
  \end{picture}
$$
Note that we have $(P_n(\Z_n(\EC_n)) \boxtimes \EC_n) \boxtimes_{\Z_n(P_n(\Z_n(\EC_n)))} P_n(\Z_n(\EC_n))=\EC_n$
}

\item There is a morphism $\rho: P_n(\Z_n(\EC_n)) \boxtimes \EC_n \to \EC_n$ defined by the gapped domain wall $P_n(\Z_n(\EC_n)) \boxtimes P_n(\Z_n(\EC_n))$ between
$\Z_n(\EC_n) \boxtimes \overline{\Z_n(\EC_n)} \boxtimes \Z_n(\EC_n)$ and $\Z_n(\EC_n)$
as depicted in the following picture:
\be \label{eq:rho}
\raisebox{-35pt}{ \begin{picture}(140, 85)
   \put(0,0){\scalebox{2}{\includegraphics{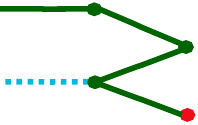}}}
   \put(0,0){
     \setlength{\unitlength}{.75pt}\put(-18,-19){
     \put(170, 75)       {\footnotesize $P_n(\Z_n(\EC_n))$}
     \put(170, 22)      {\footnotesize $\EC_n$}
     \put(50, 62)       {\footnotesize $P_n(\Z_n(\EC_n))$}
     \put(128, 49)     {\footnotesize $ \overline{\Z_n(\EC_n)} $}
     \put(97, 26)     { \footnotesize$ \Z_n(\EC_n) $}
     \put(15, 118)     {\footnotesize $ \Z_n(\EC_n) $}
     \put(70, 118)     {\footnotesize $ P_n(\Z_n(\EC_n)) $}
     \put(130, 100)   {\footnotesize $ \Z_n(\EC_n) $}
     \put(15, 38)     {\footnotesize $\one_{n+1}$ }
     }\setlength{\unitlength}{1pt}}
  \end{picture}}
\ee
Note that we have
$$
\big( P_n(\Z_n(\EC_n)) \boxtimes P_n(\Z_n(\EC_n)) \big) \boxtimes_{\Z_n(\EC_n) \boxtimes \overline{\Z_n(\EC_n)} \boxtimes \Z_n(\EC_n)}  \big( P_n(\Z_n(\EC_n)) \boxtimes \EC_n \big) = \EC_n.
$$

\enu
\end{expl}

To summarize, the collection of $n$D topological orders form a symmetric monoidal category $\TO_n$: an object is an $n$D topological order $\EC_n$, a morphism $f:\EC_n\to\ED_n$ between two objects $\EC_n,\ED_n$ is a gapped domain wall $f_n$ between $\Z_n(\EC_n)$ and $\Z_n(\ED_n)$ as defined above, and there is a symmetric binary operation $\boxtimes$ on $\TO_n$. Moreover, there exists a unique morphism $\iota_{\EC_n}:\one_n\to\EC_n$ for every object $\EC_n$ as we have seen in Example \ref{exam:mor}. These data suffice to define the universal property of the center in $\TO_n$. Before proceeding on, we would like to take a look at the special case $n=2$.

\begin{expl} \label{expl:monoidal-fun}
A 2D topological order $\EC_2$ is described mathematically by a unitary fusion category, and its \bulk $\Z_2(\EC_2)$ is given by the Drinfeld center of $\EC_2$ (see \cite{kitaev-kong,kong-anyon}). In particular, the trivial 2D topological order $\one_2$ is described by the category of finite-dimensional Hilbert spaces $\bk$.
Moreover, $P_2(\Z_2(\EC_2))$ is the underlying unitary fusion category of $\Z_2(\EC_2)$ by forgetting the braiding structure.
The stacking operation $\boxtimes$ corresponds to Deligne tensor product and, more generally, the operation $\boxtimes_{\Z_2(\EC_2)}$ is the relative tensor product (see \cite[Sec.\,5.2]{ai}). By definition, the Deligne tensor product $\EC_2\boxtimes\ED_2$ of two unitary fusion categories $\EC_2,\ED_2$ is also a unitary fusion category with the monoidal structure $(c\boxtimes d)\otimes(c'\boxtimes d') := (c\otimes c')\boxtimes(d\otimes d')$ for $c,c'\in\EC_2$ and $d,d'\in\ED_2$.

The result from \cite[Sec.\,3.2]{kong-zheng} then states that a morphism between two 2D topological orders is nothing but an (isomorphisms class of) monoidal functor.
More explicitly, if $f: \EC_2\to \ED_2$ is a unitary monoidal functor between two unitary fusion categories $\EC_2$ and $\ED_2$, then
the corresponding unitary fusion category $f_2$ is given by the category of $\EC_2$-$\ED_2$-bimodule functors $\fun_{\EC_2|\ED_2}(\ED_2, \ED_2)$. Namely, there is a canonical monoidal equivalence:
$$\fun_{\EC_2|\ED_2}(\ED_2, \ED_2)\boxtimes_{\Z_2(\EC_2)} \EC_2 \simeq \ED_2.$$
Conversely, one recovers $f$ from $f_2$ as the monoidal functor $\EC_2 \xrightarrow{1_{f_2}\boxtimes_{\Z_2(\EC_2)}\id_{\EC_2}} f_2\boxtimes_{\Z_2(\EC_2)} \EC_2 \simeq \ED_2$.
Note that the identity morphism $\id_{\EC_2}:\EC_2\to\EC_2$ is precisely the identity monoidal functor. Moreover, there exists a unique monoidal functor $\iota_{\EC_2}:\bk\to\EC_2$, the one that carries the tensor unit $\Cb$ of $\bk$ to the tensor unit $1_{\EC_2}$ of $\EC_2$.

In summary, the symmetric monoidal category $\TO_2$ is described as follows. An object is a unitary fusion category. A morphism is an (isomorphism class of) monoidal functor. The symmetric binary operation $\boxtimes$ is Deligne tensor product, and there is a distinguished object $\bk$ such that $\bk\boxtimes\EC_2\simeq\EC_2$ for all objects $\EC_2$ in $\TO_2$.
\end{expl}

\void{
The following result follows from Def.\,\ref{def:morphism-2} immediately.
\begin{thm-ph}  \label{prop:factorize}
Let $f: \EC_n \to \ED_n$ be a morphism. If $\EC_n$ is anomaly-free, then $\ED_n = \EC_n \boxtimes \EE_n$ for some $n$D topological order $\EE_n$. If $\ED_n$ is also anomaly-free, so is $\EE_n$.
\end{thm-ph}

\begin{rema} {\rm
When $n=2$, mathematically, it was known that if $f: \EA \to \EB$ is a braided monoidal functor between two modular tensor categories $\EA$ and $\EB$, then we have  $\EB\simeq \EA \boxtimes \EE$, where $\EE$ is another modular tensor category.
}
\end{rema}
}

\section{\bulk = center} \label{sec:bulk=center}

In this section, we prove that the \bulk satisfies the universal property of the center.

\medskip

Let $\EC_n$ be an $n$D topological order and let $\Z_n(\EC_n)$ be its \bulk.
First, we observe that the morphism $\rho: P_n(\Z_n(\EC_n)) \boxtimes \EC_n \to \EC_n$ from Example \ref{exam:mor} is a unital action. That is, the following diagram is commutative:
$$
\raisebox{4em}{
\xymatrix{
& P_n(\Z_n(\EC_n)) \boxtimes \EC_n \ar[rd]^\rho  & \\
\one_n\boxtimes\EC_n=\EC_n \ar[ur]^{\iota_{P_n(\Z_n(\EC_n))}\boxtimes \id_{\EC_n}}  \ar[rr]^{\id_{\EC_n}}  & & \EC_n \,\, .
}}
$$
Indeed, this follows from the following two realizations of the same physical configuration:
$$
\raisebox{-30pt}{ \begin{picture}(170, 90)
   \put(0,10){\scalebox{2}{\includegraphics{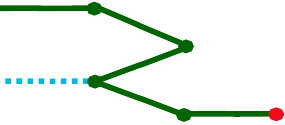}}}
   \put(0,10){
     \setlength{\unitlength}{.75pt}\put(-18,-19){
     \put(170, 75)       {\footnotesize $P_n(\Z_n(\EC_n))$}
     \put(240, 25)      {\small $\EC_n$}
     \put(50, 62)       {\footnotesize $P_n(\Z_n(\EC_n))$}
     \put(160, 10)       {\footnotesize $P_n(\Z_n(\EC_n))$}
     \put(128, 49)     {\footnotesize $ \overline{\Z_n(\EC_n)} $}
     \put(97, 26)     { \footnotesize$ \Z_n(\EC_n) $}
     \put(15, 118)     {\footnotesize $ \Z_n(\EC_n) $}
     \put(70, 118)     {\footnotesize $ P_n(\Z_n(\EC_n)) $}
     \put(130, 100)   {\footnotesize $ \Z_n(\EC_n) $}
     \put(15, 38)     {\footnotesize $\one_{n+1}$ }
     }\setlength{\unitlength}{1pt}}
  \end{picture}}
  \quad = \quad\quad
  \begin{picture}(140, 30)
   \put(0,-3){\scalebox{2}{\includegraphics{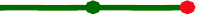}}}
   \put(-25,-3){
     \setlength{\unitlength}{.75pt}\put(-18,-70){
     \put(60,87)      {\footnotesize $\Z_n(\EC_n)$}
     \put(150, 64)     {\footnotesize $\Z_n(\EC_n)$}
     \put(200, 86)     {\footnotesize $\EC_n$}
     \put(105, 88)     {\footnotesize $P_n(\Z_n(\EC_n))$}
     }\setlength{\unitlength}{1pt}}
  \end{picture}\, ,
$$
where the left hand side depicts the morphism $\rho\circ(\iota_{P_n(\Z_n(\EC_n))}\boxtimes \id_{\EC_n})$ and the right hand side depicts the identity morphism $\id_{\EC_n}$.

\medskip

Now we are ready to state and prove the main result of this paper.
\begin{thm} \label{thm:universal}
The pair $(P_n(\Z_n(\EC_n)), \rho)$ satisfies the universal property of the center. More precisely, if $(\EX_n, f)$ is another pair, where $\EX_n$ is an $n$D topological order and $f: \EX_n \boxtimes \EC_n \to \EC_n$ is a morphism and a unital action, then there exists a unique morphism $\underline{f}: \EX_n \to P_n(\Z_n(\EC_n))$ such that the following diagram
\be  \label{diag:univ-prop-2}
\xymatrix{
& P_n(\Z_n(\EC_n)) \boxtimes \EC_n \ar[rd]^\rho  & \\
\EX_n\boxtimes\EC_n \ar[ur]^{\underline{f}\boxtimes \id_{\EC_n}}  \ar[rr]^{f}  & & \EC_n \,\,
}
\void{
\raisebox{4em}{
\xymatrix@R=1em@C=1.5em{
&  & P_n(\Z_n(\EC_n)) \boxtimes \EC_n \ar@/^1.8pc/[rrdddd]^\rho &  & \\
& & &  & \\
& & \EX_n \boxtimes \EC_n \ar[uu]^{ \underline{f} \boxtimes \id_{\EC_n}} \ar[rrdd]^f & & \\
& & & & \\
\EC_n \ar[uurr]^{\iota_{\EX_n} \boxtimes \id_{\EC_n}} \ar@/^1.8pc/[uuuurr]^{\iota_{P_n(\Z_n(\EC_n))} \boxtimes \id_{\EC_n}}  \ar[rrrr]^{\id_{\EC_n}} & & & & \EC_n\,
}}
}
\ee
is commutative.
\end{thm}
\pf
Since $f: \EX_n \boxtimes \EC_n \to \EC_n$ is a unital action, $f\circ (\iota_{\EX_n} \boxtimes \id_{\EC_n}) = \id_{\EC_n}$.
The physical configuration associated to this equality is depicted as follows:
\be \label{eq:univ-proof-3}
\raisebox{-20pt}{
 \begin{picture}(140, 60)
   \put(0,15){\scalebox{2}{\includegraphics{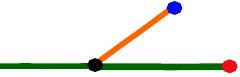}}}
   \put(0,15){
     \setlength{\unitlength}{.75pt}\put(-18,-19){
     \put(205, 32)       {\footnotesize $\EC_n$}
     \put(85, 60)       {\footnotesize $\Z_n(\EX_n)$}
     \put(80, 6)      {\footnotesize $f_n$ }
     \put(0, 10)     {\footnotesize $ \Z_n(\EC_n) $}
     \put(160, 78)     {\footnotesize $ \EX_n $}

     \put(135, 10)     {\footnotesize $\Z_n(\EC_n)$ }
     }\setlength{\unitlength}{1pt}}
  \end{picture}}
  \quad\quad = \quad\quad
   \begin{picture}(120, 30)
   \put(0,-3){\scalebox{2}{\includegraphics{pic-def-morphism-3-eps-converted-to.pdf}}}
   \put(-25,-3){
     \setlength{\unitlength}{.75pt}\put(-18,-70){
     \put(60,64)      {\footnotesize $\Z_n(\EC_n)$}
     \put(150, 64)     {\footnotesize $\Z_n(\EC_n)$}
     \put(200, 86)     {\footnotesize $\EC_n$}
     \put(105, 88)     {\footnotesize $P_n(\Z_n(\EC_n))$}
     }\setlength{\unitlength}{1pt}}
  \end{picture} ,
\ee
which implies that
\be\label{eqn:fn}
f_n \boxtimes_{\Z_n(\EX_n)} \EX_n=P_n(\Z_n(\EC_n))
\ee
as gapped domain walls between $\Z_n(\EC_n)$ and $\Z_n(\EC_n)$.
Now we regard $f_n$ as a gapped domain wall between $\Z_n(\EX_n)$ and $\Z_n(\EC_n)\boxtimes \overline{\Z_n(\EC_n)}$. According to the definition of a morphism, Eq.\,(\ref{eqn:fn}) says that $f_n$ defines a morphism $\underline{f}: \EX_n \to P_n(\Z_n(\EC_n))$.

Such defined $\underline{f}$ makes the diagram (\ref{diag:univ-prop-2}) commutative. This follows from the following two realizations of the same physical configuration:
\void{
Indeed, the commutativity of the upper-left triangle is nothing but the following identity:
$$
\raisebox{-40pt}{
 \begin{picture}(140, 90)
   \put(0,15){\scalebox{2}{\includegraphics{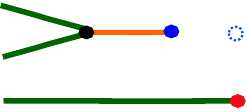}}}
   \put(0,15){
     \setlength{\unitlength}{.75pt}\put(-18,-19){
     \put(205, 32)       {\footnotesize $\EC_n$}
     \put(100, 65)       {\footnotesize $\Z_n(\EX_n)$}
     \put(82, 90)      {\footnotesize $f_n$ }
     \put(100, 33)     {\footnotesize $ \Z_n(\EC_n) $}
     \put(147, 88)     {\footnotesize $ \EX_n $}
     \put(200,88)     {\footnotesize $\one_n$}

     \put(35, 48)     {\footnotesize $\overline{\Z_n(\EC_n)}$ }
     \put(35, 103)     {\footnotesize $\Z_n(\EC_n)$ }
     }\setlength{\unitlength}{1pt}}
  \end{picture}}
  \quad\quad = \quad\quad
\raisebox{-40pt}{
   \begin{picture}(120, 30)
   \put(0,15){\scalebox{2}{\includegraphics{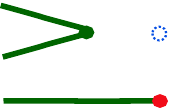}}}
   \put(0,15){
     \setlength{\unitlength}{.75pt}\put(-18,-19){
     \put(35,103)      {\footnotesize $\Z_n(\EC_n)$}
     \put(35, 48)     {\footnotesize $\overline{\Z_n(\EC_n)}$ }
     \put(145,88)     {\footnotesize $\one_n$}

     \put(100, 33)     {\footnotesize $\Z_n(\EC_n)$}
     \put(145, 32)     {\footnotesize $\EC_n$}
     \put(82, 90)     {\footnotesize $P_n(\Z_n(\EC_n))$}
     }\setlength{\unitlength}{1pt}}
  \end{picture}} ,
$$
and that of the upper-right triangle is nothing but the following identity:
}
\be \label{eq:ff}
\raisebox{-35pt}{ \begin{picture}(170, 90)
   \put(0,10){\scalebox{2}{\includegraphics{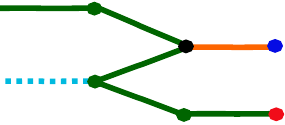}}}
   \put(0,10){
     \setlength{\unitlength}{.75pt}\put(-18,-19){
     \put(165, 88)       {\footnotesize $f_n$}
     \put(240,78)       {\footnotesize $\EX_n$}
     \put(180,66)       {\footnotesize $\Z_n(\EX_n)$}
     \put(180,36)       {\footnotesize $\Z_n(\EC_n)$}

     \put(240, 25)      {\small $\EC_n$}
     \put(50, 62)       {\footnotesize $P_n(\Z_n(\EC_n))$}
     \put(160, 10)       {\footnotesize $P_n(\Z_n(\EC_n))$}
     \put(128, 49)     {\footnotesize $ \overline{\Z_n(\EC_n)} $}
     \put(97, 26)     { \footnotesize$ \Z_n(\EC_n) $}
     \put(15, 118)     {\footnotesize $ \Z_n(\EC_n) $}
     \put(70, 118)     {\footnotesize $ P_n(\Z_n(\EC_n)) $}
     \put(130, 100)   {\footnotesize $ \Z_n(\EC_n) $}
     \put(15, 38)     {\footnotesize $\one_{n+1}$ }
     }\setlength{\unitlength}{1pt}}
  \end{picture}}
  \quad = \quad\quad
  \raisebox{-18pt}{\begin{picture}(130, 30)
   \put(0,-3){\scalebox{2}{\includegraphics{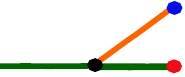}}}
   \put(-25,-3){
     \setlength{\unitlength}{.75pt}\put(-18,-70){
     \put(60,87)      {\footnotesize $\Z_n(\EC_n)$}
     \put(140, 64)     {\footnotesize $\Z_n(\EC_n)$}
     \put(195, 76)     {\footnotesize $\EC_n$}
     \put(195, 120)     {\footnotesize $\EX_n$}
     \put(115, 90)     {\footnotesize $f_n$}
     }\setlength{\unitlength}{1pt}}
  \end{picture}}\, ,
\ee
where the left hand side depicts the morphism $\rho\circ(\underline{f} \boxtimes \id_{\EC_n})$ and the right hand side depicts the morphism $f$.

The uniqueness of $\underline{f}$ also follows from above equality. More precisely, if $g: \EX_n \to  P_n(\Z_n(\EC_n))$ is another morphism making the diagram (\ref{diag:univ-prop-2}) commutative, then the equality (\ref{eq:ff}), with the $f_n$ on the left hand side replaced by $g_n$, holds. This new identity immediately implies that $g_n=f_n$, i.e. $g=\underline{f}$.
\epf

Note that the universal property determines the pair $(P_n(\Z_n(\EC_n)), \rho)$ up to canonical isomorphism. That is, if $(\EY_n, \gamma)$ is another pair satisfying the universal property of the center, then the morphism $\underline{\gamma}: \EY_n \to P_n(\Z_n(\EC_n))$ induced by $\gamma$ is inverse to the morphism $\underline{\rho}: P_n(\Z_n(\EC_n)) \to \EY_n$ induced by $\rho$ hence identifies $\EY_n$ with $P_n(\Z_n(\EC_n))$. In another word, the pair $(P_n(\Z_n(\EC_n)), \rho)$ is determined by the universal property without ambiguity. Under the terminology of mathematics, the pair $(P_n(\Z_n(\EC_n)), \rho)$ is the {\it center} of the $n$D topological order $\EC_n$.

\medskip

Recall that the $n$D topological order $P_n(\Z_n(\EC_n))$ can be obtained by double folding the $n+$1D topological order $\Z_n(\EC_n)$. To recover $\Z_n(\EC_n)$ from $P_n(\Z_n(\EC_n))$, it suffices to reverse the double folding process, i.e. to split $\Z_n(P_n(\Z_n(\EC_n)))$. This is possible if we use the additional data $\rho$, because $\rho_n$ is an invertible domain wall between $\Z_n(P_n(\Z_n(\EC_n)))$ and $\Z_n(\EC_n) \boxtimes \overline{\Z_n(\EC_n)}$ hence identifies them (recall the discussions associated to Fig.\,\ref{fig:tunneling}). Namely, $\rho$ provides a splitting of $\Z_n(P_n(\Z_n(\EC_n)))$ thus recovers $\Z_n(\EC_n)$. This shows that the pair $(P_n(\Z_n(\EC_n)),\rho)$ contains the same information as $\Z_n(\EC_n)$.

We reach the conclusion  ``\bulk = center''. This result is independent of how we describe the boundary phase and the \bulk phase mathematically.



\begin{expl} {\rm
Let us proceed on Example \ref{expl:monoidal-fun}.
The morphism $\rho: P_2(\Z_2(\EC_2)) \boxtimes \EC_2 \to \EC_2$ is given by the  monoidal functor
$$P_2(\Z_2(\EC_2)) \boxtimes \EC_2 \to (P_2(\Z_2(\EC_2)) \boxtimes P_2(\Z_2(\EC_2))) \boxtimes_{\Z_2(\EC_2) \boxtimes \overline{\Z_2(\EC_2)} \boxtimes \Z_2(\EC_2)} (P_2(\Z_2(\EC_2)) \boxtimes \EC_2)  \simeq \EC_2,
$$
or equivalently, by the composed functor $P_2(\Z_2(\EC_2)) \boxtimes \EC_2 \to \EC_2 \boxtimes\EC_2 \xrightarrow{\otimes} \EC_2$. It is clear that $\rho$ is a unital action, i.e. $\rho(1_{P_2(\Z_2(\EC_2))}\boxtimes a)\simeq a$ for $a\in\EC_2$.

Let $f: \EX_2 \boxtimes \EC_2 \to \EC_2$ be a monoidal functor which is also a unital action. In particular, $f$ is equipped with natural isomorphisms $f(x\boxtimes a)\otimes f(y\boxtimes b) \simeq f((x\otimes y)\boxtimes (a\otimes b))$ and $f(1_{\EX_2}\boxtimes a) \simeq a$.
Then there is a monoidal functor $\underline{f}: \EX_2 \to \EC_2$ defined by $x \mapsto f(x\boxtimes 1_{\EC_2})$. Note that the object $f(x\boxtimes 1_{\EC_2})$ in $\EC_2$ is naturally equipped with a half-braiding
$$
f(x\boxtimes 1_{\EC_2}) \otimes a \simeq
f(x\boxtimes 1_{\EC_2}) \otimes f(1_{\EX_2} \boxtimes a)
\simeq f(x\boxtimes a) \simeq
f(1_{\EX_2} \boxtimes a) \otimes f(x\boxtimes 1_{\EC_2})
\simeq a \otimes f(x\boxtimes 1_{\EC_2}).
$$
In other words, $\underline{f}$ defines a monoidal functor from $\EX_2$ to $P_2(\Z_2(\EC_2))$. Moreover, the following diagram
$$
\xymatrix{
& P_2(\Z_2(\EC_2)) \boxtimes \EC_2 \ar[rd]^\rho  & \\
\EX_2\boxtimes \EC_2 \ar[rr]^{f} \ar[ru]^{\underline{f} \boxtimes \id_{\EC_2}}  & & \EC_2
}
$$
is commutative. The uniqueness of $\underline{f}$ is also easy to see. This shows that the pair $(P_2(\Z_2(\EC_2)), \rho)$ satisfies the universal property of the center.

Recall that $P_2(\Z_2(\EC_2))$ is obtained from $\Z_2(\EC_2)$ by forgetting the braiding structure. As we have argued, one can use $\rho$ to recover $\Z_2(\EC_2)$ from $P_2(\Z_2(\EC_2))$. Indeed, by the universal property of the center, the composed monoidal functor $\mu: P_2(\Z_2(\EC_2)) \boxtimes P_2(\Z_2(\EC_2)) \boxtimes \EC_2 \xrightarrow{\id\boxtimes\rho} P_2(\Z_2(\EC_2)) \boxtimes \EC_2  \xrightarrow{\rho} \EC_2$ determines a monoidal functor $\underline{\mu}: P_2(\Z_2(\EC_2)) \boxtimes P_2(\Z_2(\EC_2)) \to P_2(\Z_2(\EC_2))$. The uniqueness of $\underline{\mu}$ forces it to be the obvious one, the tensor product functor $\otimes$, and the monoidalness of $\underline{\mu}$ supplies a natural isomorphism $\underline{\mu}(a\boxtimes b) \otimes \underline{\mu}(c\boxtimes d) \simeq \underline{\mu}((a\otimes c)\boxtimes(b\otimes d))$ for $a,b,c,d\in P_2(\Z_2(\EC_2))$. Then the forgotten braiding structure on $P_2(\Z_2(\EC_2))$ can be recovered by the following natural isomorphisms $b\otimes c \simeq \underline{\mu}(1\boxtimes b) \otimes \underline{\mu}(c\boxtimes1) \simeq \underline{\mu}((1\otimes c)\boxtimes(b\otimes1)) \simeq \underline{\mu}(c\boxtimes b) \simeq c\otimes b$.
}
\end{expl}

\begin{expl} {\rm
The only anomaly-free 1+1D topological order is the trivial one $\one_2$, which is described by the category of finite-dimensional Hilbert spaces $\bk$. It was explained in \cite[Example 2.25]{kong-wen-zheng} that the only 1D topological orders can be described by a pair $(\bk, u)$,
where $u$ is an object of $\bk$. Moreover, the stacking operation is given by $(\bk, u) \boxtimes (\bk, v) = (\bk, u\otimes v)$. Note that $\Z_1(\bk, u) = \bk$, $P_1(\Z_1(\bk, u)) = \one_1 = (\bk,\mathbb{C})$ and $\rho: P_1(\Z_1(\bk, u)) \boxtimes (\bk, u) \to (\bk, u)$ is the identity morphism of $(\bk,u)$. That is, both of $\Z_1(\bk, u)$ and $(P_1(\Z_1(\bk, u)) ,\rho)$ are trivial.
By definition, a morphism $(\bk, u) \to (\bk, u')$ is a 1D topological order $(\bk, v)$ such that $(\bk, v) \boxtimes (\bk, u) = (\bk, u')$; this is equivalent to a functor $\bk \to \bk$ that carries $u$ to $u'$ (and carries $\mathbb{C}$ to $v$).
The pair $(P_1(\Z_1(\bk, u)) ,\rho)$ satisfies the universal property of the center trivially: if $\EX_1\boxtimes (\bk, u) \to (\bk, u)$ is a morphism then $\EX_1 = (\bk,\mathbb{C})$ which is identical to $P_1(\Z_1(\bk, u))$.
This ``\bulk = center'' relation still holds even if we include unstable 1+1D phases, which occur naturally in dimensional reduction processes (see \cite[Example 3.7,\,6.4]{kong-wen-zheng} and also \cite[Sec.\,5.2]{ai}).
}
\end{expl}

\begin{rema} \label{rema:prediction}
Theorem \ref{thm:universal} is a non-trivial result, which gives concrete physical predictions. For example, in the 3+1D Walker-Wang model \cite{ww} built on a unitary pre-modular tensor 1-category $\EC$, which can be viewed as a monoidal 2-category, Theorem \ref{thm:universal} implies that the topological excitations in the 3+1D bulk shall form the braided monoidal 2-category \cite{kv,ds} given by the monoidal center of the monoidal 2-category $\EC$ constructed by Baez and Neuchl \cite{baez}. We believe that Walker-Wang's construction can be generalized to 3+1D lattice models based on a generic unitary fusion 2-category $\EC$ \cite{kong-wen-zheng}, and the bulk excitations in this conjectural model should form a unitary braided fusion 2-category given by the monoidal center of $\EC$. Moreover, ``\bulk = center'' provides a serious constraint to the precise mathematical formulation of topological orders in all dimensions. It led us to propose in \cite{kong-wen-zheng} a categorical description of the topological excitations in potentially anomalous topological orders in all dimensions.
\end{rema}

\begin{rema}
It was explained in \cite{kong-wen-zheng} that ``\bulk = center'' discussed in this work is only the first layer of the complete boundary-bulk relation, which can be summarized as the functoriality of the center. For 2+1D topological orders with gapped boundaries, this functoriality of Drinfeld center of unitary fusion categories was proved rigorously in \cite{kong-zheng}.
\end{rema}

\begin{rema}
Our main result also sheds lights on a similar boundary-bulk duality (i.e. open-closed duality) in 2D rational CFT's \cite{ffrs,kr,dav,bklr}. It will also be interesting to study its relation to some results in factorization algebras (see \cite{lurie,francis,aft,ginot,bbj2,ai}).
\end{rema}

Before we conclude this paper, we would like to give another important remark, which has already led to some new results \cite{kz2,zheng,kz3}.

\begin{rema}
If an $n+$1D anomaly-free topological order has a topologically protected gapless $n$D boundary phase and if the unique-bulk hypothesis still holds, then it is easy to see that our proof also works for the gapless boundary case. Therefore, the gapped bulk phase associated to a gapless boundary phase should also be given by the center of the boundary phase. In particular, in 2+1D quantum Hall systems, this result suggests that there should be a mathematical description of the 1+1D gapless edge modes such that its center gives the modular tensor category $\EC$ that describes the 2+1D bulk phase. In \cite{kz2}, it was shown that there is an enriched monoidal category $\EC^\sharp$ (enriched by boundary CFT's) such that its Drinfeld center is exactly $\EC$. In \cite{kz3}, we showed that such enriched monoidal categories give a mathematical description of the gapless boundary phases of 2+1D topological orders. Such a mathematical description also provides a way to extend the Reshetikhin-Turaev 2+1D TQFT down to points \cite{zheng}.
\end{rema}

\end{document}